\documentclass[conference]{IEEEtran}
\usepackage{booktabs}
\usepackage{xcolor}
\usepackage{colortbl}
\usepackage{amssymb}
\usepackage{algorithmic}
\usepackage{amsmath} 
\usepackage{amsfonts}

\ifCLASSINFOpdf
\usepackage[pdftex]{graphicx}
\else
\usepackage[dvips]{graphicx}
\fi
\usepackage{epstopdf}
\usepackage{array}
\ifCLASSOPTIONcompsoc
\usepackage[caption=false,font=normalsize,labelfont=sf,textfont=sf]{subfig}
\else
\usepackage[caption=false,font=footnotesize]{subfig}
\fi
\usepackage{float}
\usepackage{fixltx2e}
\usepackage{stfloats}
\usepackage{color}
\usepackage{xcolor}
\usepackage{textcomp}
\usepackage[normalem]{ulem}
\usepackage{cite}
\usepackage{multibib} 
\usepackage{caption}
\usepackage{nomencl}
\makenomenclature
\usepackage{etoolbox}
\renewcommand\nomgroup[1]{%
	\item[\bfseries
	\ifstrequal{#1}{P}{Physical Variables}{%
		\ifstrequal{#1}{M}{Mathematical Symbols}{%
			\ifstrequal{#1}{S}{Subscripts and Superscripts}{}}} ]}

\usepackage{hyperref}
\hypersetup{
	colorlinks=true,
	linkcolor=blue,
	filecolor=magenta,      
	urlcolor=cyan,}

\makeatletter
\newcommand*\titleheader[1]{\gdef\@titleheader{#1}}
\AtBeginDocument{%
	\let\st@red@title\@title%
	\def\@title{%
		\bgroup\normalfont\large\centering\@titleheader\par\egroup
		\vskip0.5em\st@red@title}
}
\makeatother

\title{Synthesizing Distributed Energy Resources in Microgrids with Temporal Logic Specifications}
\titleheader{2018 IEEE International Symposium on Power Electronics for Distributed Generation Systems (PEDG), Charlotte, USA}
\IEEEoverridecommandlockouts

\begin{document}
\author{
\IEEEauthorblockN{Yichen Zhang\IEEEauthorrefmark{1},
		Mohammed Olama\IEEEauthorrefmark{2},
		Alexander Melin\IEEEauthorrefmark{2},
		Yaosuo Xue\IEEEauthorrefmark{2},
		Seddik Djouadi\IEEEauthorrefmark{1},
		Kevin Tomsovic\IEEEauthorrefmark{1}}
\IEEEauthorblockA{\IEEEauthorrefmark{1}Department of Electrical Engineering and Computer Science, University of Tennessee, Knoxville, TN 37996}
\IEEEauthorblockA{Email: {\tt\small \{yzhan124, mdjouadi, tomsovic\}@utk.edu.}}
\IEEEauthorblockA{\IEEEauthorrefmark{2} Oak Ridge National Laboratory, Oak Ridge, TN 37831}
\IEEEauthorblockA{Email: {\tt\small \{olamahussemm, melina, xuey@ornl.gov\}@ornl.gov.}}
\thanks{This manuscript has been authored by UT-Battelle, LLC, under contract DE-AC05-00OR22725 with the U.S. Dept. of Energy.}}

\maketitle

\begin{abstract}
Grid supportive (GS) modes integrated within distributed energy resources (DERs) can improve the frequency response. However, synthesis of GS modes for guaranteed performance is challenging. Moreover, a tool is needed to handle sophisticated specifications from grid codes and protection relays. This paper proposes a model predictive control (MPC)-based mode synthesis methodology, which can accommodate the temporal logic specifications (TLSs). The TLSs allow richer descriptions of control specifications addressing both magnitude and time at the same time. The proposed controller will compute a series of Boolean control signals to synthesize the GS mode of DERs by solving the MPC problem under the normal condition, where the frequency response predicted by a reduced-order model satisfies the defined specifications. Once a sizable disturbance is detected, the pre-calculated signals are applied to the DERs. The proposed synthesis methodology is verified on the full nonlinear model in Simulink. A robust factor is imposed on the specifications to compensate the response mismatch between the reduce-order model and nonlinear model so that the nonlinear response satisfies the required TLS.
\end{abstract}

\IEEEpeerreviewmaketitle

\section{Introduction}\label{sec_introd}

Microgrids have become an ideal solution for powering remote locations \cite{allen2016sustainable}. Such microgrids are usually fed by mixed sources of diesel generators (DGs) and distributed energy resources (DERs) to reduce the cost \cite{allen2016sustainable}. Most DERs are converter-interfaced and do not respond to frequency excursions in the grid. Such characteristics of DERs can severely degrade the performance of frequency response with increasing penetration, leading to larger rate of change of frequency and frequency excursion during the transient period \cite{Pulgar2018Inertia}. This could lead to a potential trip of any rotating machine connected to the network, or possibly trigger unnecessary frequency relays, in which case the system has adequate capacity to attain a stable steady-state \cite{freq_limit_storage}. Currently, lots of grid supportive (GS) modes have been integrated into the converter active power control loops. Among all converter-interfaced DERs, the wind turbine generators (WTGs) are preferred to be integrated with supportive functions due to the large amount of available kinetic energy.

On the one hand, synthesis of such hybrid controllers to achieve certain system performance specifications is a challenging task since most GS modes are feeding with only local information and lack of grid dynamics awareness capacity. This issue are tackled from two aspects. Refs. \cite{zyc_hybrid_controller,rate_MIT,freq_limit_storage,maram2014event} employ certain simplified models to estimate the frequency dynamics, based on which the control are designed. Alternatively, certain pre-event simulations are carried out to determine the commitment of GS modes in  \cite{freq_limit_load,Wang2018Quantitative}. 

On the other hand, performance specifications in most literature are only state-dependent. But the protection replays of power system in real industry are designed based on states and dwell time simultaneously. Most of the grid codes also allow states to enter certain restricted regions, but the dwell time should not be larger than a specified value. So, it is natural to seek a tool that can specify time and region requirements in control designs. The temporal logic specifications (TLSs) allow richer descriptions of specifications including set, logic and time-related properties. For example, to guarantee the proper operation of microgrids, the speed deviation of the synchronous generator should not exceed $\pm1.5$ Hz for 0.1 second \cite{xyngi2010smart}. The pioneering work in \cite{xu2017energy} introduces the TLSs for controller synthesis of energy storage systems, where the frequency is requried to restore back to $60\pm 0.2$ Hz within 2 seconds. 

In this paper, inspired by both \cite{zyc_hybrid_controller} and \cite{freq_limit_load} and motived by the introduction of TLSs \cite{xu2017energy}, a model predictive control (MPC)-based control synthesis methodology is proposed, which can accommodate the TLSs. The controller is configured into two levels including the scheduling level and the triggering level. In the scheduling level, a series of Boolean control signals are computed by solving the MPC problem, where the frequency response predicted by a reduced-order model satisfies the defined specifications under a given worse-case contingency. In addition, the scheduling level will constantly re-schedule the signal based on the operating condition and varying specifications. The triggering level will measure the frequency and detect whether a severe contingency close to the worst case is happening. Once such a contingency is detected, the scheduled signals are applied to the WTGs. The performance will be guaranteed by the scheduling level if the analytical model can precisely estimate the system behavior. The overall configuration is analogous to the adaptive remedial action scheme in \cite{maram2014event}.

The reminder of the paper is organized as follows. Section \ref{sec_pre} introduces preliminary knowledge about TLSs. Section \ref{sec_model} introduces the models of microgrids with DGs and WTGs, where the analytical models are derived. Section \ref{sec_method} introduces the MPC-based control synthesis methodology, including the overall configuration, MPC formulation for the scheduling level, and the results with nonlinear simulation verifications. Conclusions and future works are discussed in Section \ref{sec_conclusion}.

\section{Preliminaries on Temporal Logic Specification}\label{sec_pre}
A temporal logic specification is built by combining the atomic propositions (APs) using logical and temporal operators. An AP is a statement on the system variables that is either true or false for some given value of the systems variables \cite{karaman2008optimal}. For example, the statement \textquotedblleft the grid frequency deviation should never exceed ±0.5 Hz" is an AP. The commonly used logical operators are conjunction ($\wedge$), disjunction ($\vee$), negation ($\neg$), implication ($\rightarrow$), and equivalence ($\leftrightarrow$). The temporal operators include eventually ($\lozenge$), always ($\square$), and until ($U$). The TLSs can be categorized into two groups, that is, discrete-time and continuous-time TLSs. For a discrete-time TLS, timing intervals cannot be added with the temporal operators. For example, $\lozenge p$ for $p=(y<5)$ states the output $y$ will be eventually less than five without specifying when the condition will be fulfilled. As a supplementary, a continuous-time TLS can add the timing intervals like $\lozenge_{[2,+∞]}p$ for $p=(y<5)$, which states the output $y$ will be eventually less than five after two seconds. In this paper, the continuous-time TLSs are employed. 

Considerable efforts have been devoted to control synthesis for continuous-time TLSs. On the one hand, in \cite{xu2017energy,winn2015safety}, the temporal logic constraints are substituted into the optimization objectives, leading to a unconstrained problem that can be solved by some functional gradient descent algorithms. On the other hand, the authors in \cite{Raman2014} introduce an approach using mixed-integer convex optimization to encode the TLSs as constraints. First, the safe or unsafe sets are represented as polyhedrons (by finite many hyperplanes). An AP like $x\in P$ can be formulated as a linear program. Second, some integer variables are introduced to indicate whether the condition holds or not. The if and else condition can be formulated in the linear program using the big-M technique. Finally, the overall problem becomes a mix-integer linear program (MILP). The encoding procedure has been implemented in the toolbox BluSTL \cite{donze2015blustl}, which is employed for problem conversion here. The detailed procedure of encoding TLSs into MILP is out of scope of this paper.

\section{Microgrid Modeling with DGs and WTGs}\label{sec_model}
\begin{figure}[t]
	\centering
	\includegraphics[scale=0.42]{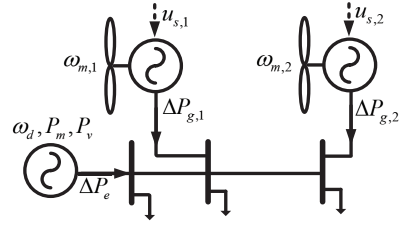}
	\vspace{-0.15in}
	\caption{Studied microgrid consisting of one DG and two WTGs.}
	\label{fig_Model_Network}
\end{figure}
\begin{figure*}[t]
	\centering
	\includegraphics[scale=0.3]{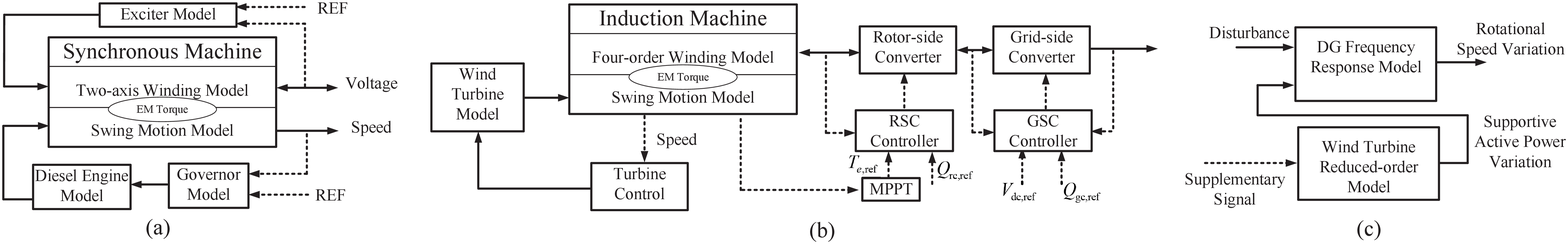}
	\vspace{-0.15in}
	\caption{(a) Modules and their interactions of diesel generation. (b) Modules and their interactions of wind turbine generator. (c) Augmented frequency response model.}
	\label{fig_Model_Full}
\end{figure*}
In this paper, a microgrid consisting of one DG and two WTGs illustrated in Fig. \ref{fig_Model_Network} is considered. The modules of DG and WTG are shown in Fig. \ref{fig_Model_Full} (a) and (b), respectively, and implemented in Simulink. The main objective of this section is to derive the augmented frequency response (AFR) model illustrated in Fig. \ref{fig_Model_Full} (c), which describes the microgrid frequency dynamics (considered equivalent to the speed of DG) subjected to both disturbances and supports. Such models have been shown to be crucial for frequency studies \cite{Shi2018Analytical}, especially the input-output relation of WTGs from the GS signal to the virtual angle \cite{hu2017modeling} and active power variation \cite{zyc_MRC_TPS_sub}. In this paper, the analytical model of AFR derived in \cite{zyc_MRC_TPS_sub} is employed. For clear illustration, the derivations are briefly repeated in this section with necessary modifications. Definitions of well-known variables can be referred to \cite{zyc_MRC_TPS_sub} and will be omitted here due to space limitations.

\subsection{Diesel Generator and Its Analytical Model}
A diesel generator (DG) is a combustion engine driven synchronous generator. A complete model consisting of a two-axis synchronous machine, combustion engine, governor, and exciter shown in Fig. \ref{fig_Model_Full} (a) is implemented in Simulink for simulation verification in Section \ref{sec_sub_sim}. The governor, engine, and swing dynamics shown in (\ref{eq_SFR}) are extracted to describe the frequency characteristics of the diesel generator, which has proved to be precise in many power system applications \cite{anderson1990low}
\begin{align}
\label{eq_SFR}
\begin{aligned}
2H_{d}\Delta\dot{\omega}_{d}&=\overline{f}(\Delta P_{m}-\Delta P_{e})\\
\tau_{d}\Delta\dot{P}_{m}&=-\Delta P_{m}+\Delta P_{v}\\
\tau_{g}\Delta\dot{P}_{v}&= -\Delta P_{v}  - \Delta\omega_{d}/(\overline{f}R_{D})
\end{aligned}
\end{align}
where $w_{d}$, $P_{m}$, $P_{v}$ are rotating speed, mechanical power, and valve position, respectively.

\subsection{Double Fed Induction Generator (DFIG)-Based WTG and Its Analytical Model}
The full nonlinear model of DFIG-based WTG is illustrated in Fig. \ref{fig_Model_Full} (b). To derive its analytical model, two steps are needed. First, the relevant modules within the WTG are selected and their mathematical models are derived. Second, the selective modal analysis (SMA)-based model reduction method is applied to obtain a first-oder model \cite{hector}, expressing the input-output relation from the GS signal to the active power variation.

In the time scale of dynamic frequency response, the most relevant modules in a WTG are the induction machine and its speed regulator, that is, the rotor-side converter (RSC) control. The dynamics of grid-side converter (GSC) are usually ten times faster than that of RSC current loop for stability reasons \cite{ying2018impact}, and thus the GSC and corresponding controller can be omitted.\par

\begin{figure}[htbp!]
	\centering
	\includegraphics[scale=0.38]{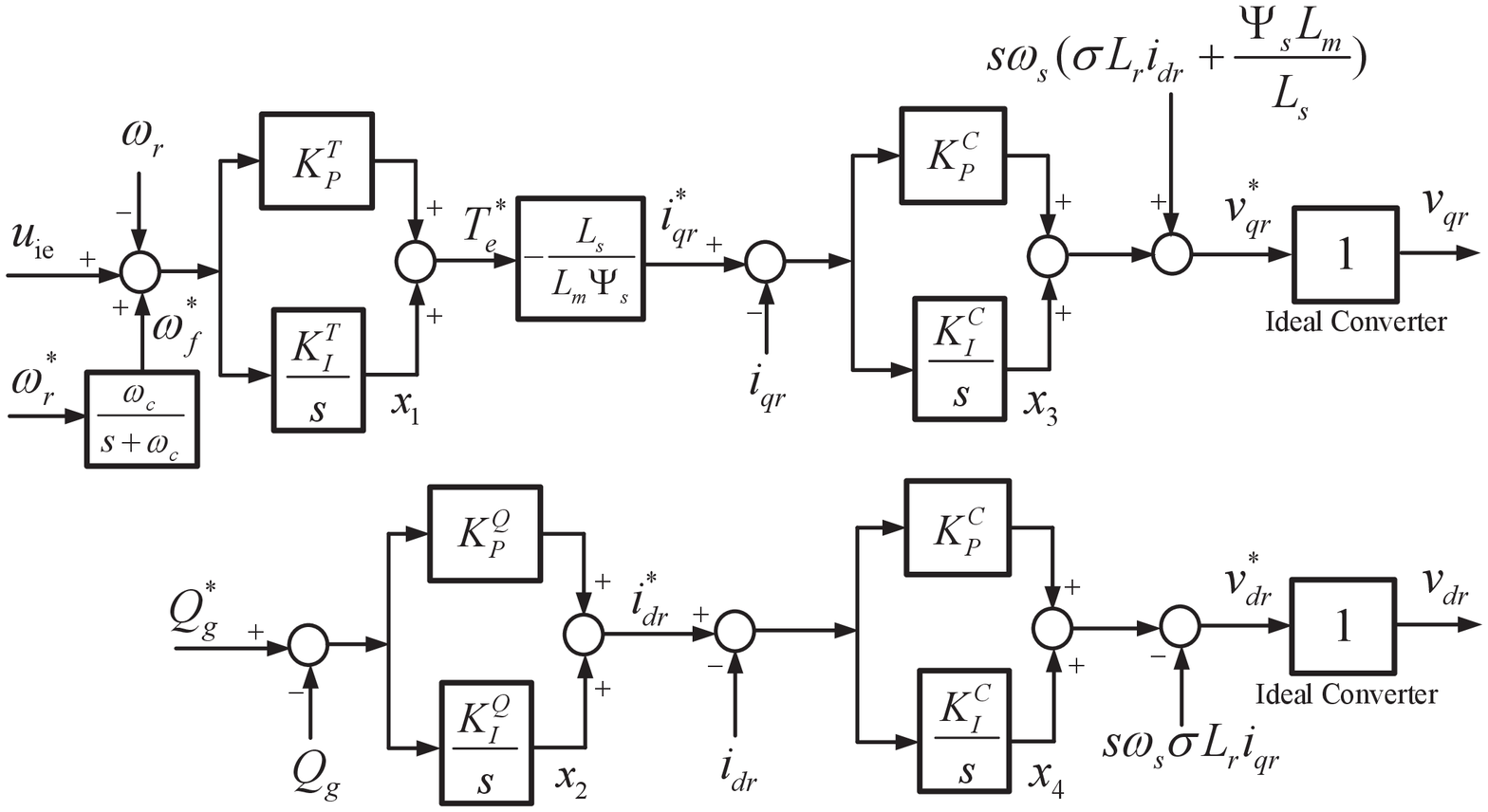}
	\vspace{-0.15in}
	\caption{Field-oriented rotor-side converter control to regulate the rotor speed of the DFIG-based WTG.}
	\label{fig_Control_Rotor}
\end{figure}

The complete RSC controller is illustrated in Fig. \ref{fig_Control_Rotor}, where the output of each integrator is defined as a state of the system. Then, the most relevant modules with respect to frequency control in a WTG are defined by the following set of differential-algebraic equations
\begin{align}
&\dot{\psi}_{qs}= \overline{\omega}(v_{qs} - R_{s}i_{qs} - \omega_{s}\psi_{ds})\label{eq_IM_ode1}\\
&\dot{\psi}_{ds} = \overline{\omega}(v_{ds} - R_{s}i_{ds} + \omega_{s}\psi_{qs})\label{eq_IM_ode2}\\
&\dot{\psi}_{qr} = \overline{\omega}[v_{qr} - R_{r}i_{qr} - (\omega_{s}-\omega_{r})\psi_{dr}]\label{eq_IM_ode3}\\
&\dot{\psi}_{dr} = \overline{\omega}[v_{dr} - R_{r}i_{dr} + (\omega_{s}-\omega_{r})\psi_{qr}]\label{eq_IM_ode4}\\
&\dot{\omega}_{r}=1/(2H_{T})(T_{m}-T_{e})\label{eq_IM_ode5}\\
&\dot{\omega}_{f}^{*}=\omega_{c}(\omega^{*}_{r}-\omega_{f}^{*})\label{eq_RSC_control_ode1}\\
&\dot{x}_{1}=K_{I}^{T}(\omega_{f}^{*}-\omega_{r}+u_{\text{ie}})\label{eq_RSC_control_ode2}\\
&\dot{x}_{2}=K_{I}^{Q}(Q^{*}_{g}-Q_{g})\label{eq_RSC_control_ode3}\\
&\dot{x}_{3}=K_{I}^{C}(i_{qr}^{*}-i_{qr})\label{eq_RSC_control_ode4}\\
&\dot{x}_{4}=K_{I}^{C}(i_{dr}^{*}-i_{dr})\label{eq_RSC_control_ode5}\\
&0=-\psi_{qs}+L_{s}i_{qs} + L_{m}i_{qr}\label{eq_IM_alg1}\\
&0=-\psi_{ds}+L_{s}i_{ds} + L_{m}i_{dr}\label{eq_IM_alg2}\\
&0=-\psi_{qr}+L_{r}i_{qr} + L_{m}i_{qs}\label{eq_IM_alg3}\\
&0=-\psi_{dr}+L_{r}i_{dr} + L_{m}i_{ds}\label{eq_IM_alg4}\\
&0=P_{g}+(v_{qs}i_{qs}+v_{qs}i_{qs}) + (v_{qr}i_{qr}+v_{qr}i_{qr})\label{eq_IM_alg5}\\
&0=Q_{g}+(v_{qs}i_{ds}-v_{ds}i_{qs}) + (v_{qr}i_{dr}-v_{dr}i_{qr})\label{eq_IM_alg6}\\
&\begin{aligned}\label{eq_RSC_control_alg1}
&0=-v_{qr}+x_{3}+K_{P}^{C}(i_{qr}^{*}-i_{qr})\\
&\qquad\qquad\qquad+(\omega_{s}-\omega_{r})(\sigma L_{r}i_{dr}+\frac{\Psi_{s}L_{m}}{L_{s}})
\end{aligned}\\
&\begin{aligned}\label{eq_RSC_control_alg2}
&0=-v_{dr}+x_{4}+K_{P}^{C}(i_{dr}^{*}-i_{dr})\\
&\qquad\qquad\qquad-(\omega_{s}-\omega_{r})\sigma L_{r}i_{qr}
\end{aligned}
\end{align}\par

Eq. (\ref{eq_IM_ode1})-(\ref{eq_IM_ode5}) are the dynamics of the induction machine in the synchronous $dq$ reference frame \cite{krause2013}, where $T_{m}$ is the mechanical torque in per unit and can be calculated according to the widely-used wind turbine model in \cite{hector}. The electromagnetic torque reads
\begin{equation}
\label{eq_IM_torque}
\begin{aligned}
T_{e}=\frac{L_{m}}{L_{s}}(\psi_{qs}i_{dr}-\psi_{ds}i_{qr})
\end{aligned}
\end{equation}
The algebraic relations of flux linkages and electric power are expressed in (\ref{eq_IM_alg1})-(\ref{eq_IM_alg6}), where $L_{s}=L_{ls}+L_{m}$ and $L_{r}=L_{lr}+L_{m}$. All values are in per unit. The rotor-side variables have been appropriately transferred to the stator side.

The dynamic model of the RSC control is given in (\ref{eq_RSC_control_ode1})-(\ref{eq_RSC_control_ode5}). The optimal speed is obtained from the maximum power point curve approximated by the following polynomial \cite{fundamental2016}
\begin{align}\label{eq_MPPT}
\omega^{*}_{r}=-0.67\times(\eta P_{g})^{2}+1.42\times(\eta P_{g})+0.51
\end{align}
for $\omega_r\in[0.8, 1.2]$. The variable $\eta$ is the ratio between the base of the induction machine and wind turbine.  Other intermediate variables are given as
\begin{equation}\label{eq_inter_control}
\begin{aligned}
& i_{qr}^{*}=\frac{-L_{s}T_{e}^{*}}{L_{m}\Psi_{s}}=\frac{-L_{s}}{L_{m}\Psi_{s}}[x_{1}+K_{P}^{T}(\omega^{*}_{f}-\omega_{r}+u_{\text{ie}})]\\
& i_{dr}^{*}=x_{2}+K_{P}^{Q}(Q^{*}_{g}-Q_{g})
\end{aligned}
\end{equation}
The time scale of converter regulation compared to the frequency response is small enough to be neglected such that $v_{qr}=v_{qr}^{*}$ and $v_{dr}=v_{dr}^{*}$. Then, the loop is closed by the algebraic relations in (\ref{eq_RSC_control_alg1})-(\ref{eq_RSC_control_alg2}), where $\sigma L_{r}=L_{r} - (L_{m}^{2})/L_{s}$. The variables $u_{\text{ie}}$ and $Q^{*}_{g}$ are control inputs while $v_{ds}$ and $v_{qs}$ are terminal conditions.

To reach the AFR in Fig. \ref{fig_Model_Full} (c), the derivation of the selective modal analysis (SMA)-based model reduction in \cite{hector,zyc_MRC_TPS_sub,zyc_hybrid_controller} is expressed. Define the state vector as
\begin{equation}\label{eq_linear_states}
\begin{aligned}
x_{w}=\left[\psi_{qs},\psi_{ds},\psi_{qr},\psi_{dr},\omega_{r},\omega^{*}_{f},x_{1},x_{2},x_{3},x_{4}\right]^{T}
\end{aligned}
\end{equation} 
Linearizing Eqs. (\ref{eq_IM_ode1})-(\ref{eq_RSC_control_alg2}) about the equilibrium point given in Section \ref{sec_sub_sim} yields the state-space model as follows
\begin{equation}\label{eq_linear_ss_full}
\begin{aligned}
\Delta\dot{x}_{w}&=A_{\text{sys}}\Delta x_{w} + B_{\text{sys}}u_{s}\\
\Delta P_{g}&=C_{\text{sys}}\Delta x_{w} +  D_{\text{sys}}u_{s}
\end{aligned}
\end{equation}
where $\Delta P_{g}$ is the active power variation of a WTG due to the GS signal $u_{s}$. The dynamics of the WTG rotor speed $\Delta\omega_{r}$ is considered as the most relevant state, while the other states denoted as $z(t)$ are less relevant. The most relevant dynamic is described by \cite{hector}
\begin{equation}
\label{eq_linear_ss_more}
\Delta\dot{\omega}_{r}=A_{11}\Delta\omega_{r}+A_{12}z + B_{r}u_{s}
\end{equation}
while the less relevant dynamics are
\begin{equation}
\label{eq_linear_ss_less}
\dot{z}=A_{22}z+A_{21}\Delta\omega_{r} + B_{z}u_{s}
\end{equation}
and the output is
\begin{equation}
\label{eq_linear_ss_out}
\Delta P_{g}=C_{r}\Delta\omega_{r}+C_{z}z+D_{\text{sys}}u_{s}
\end{equation}
The solution of (\ref{eq_linear_ss_less}) can be represented as
\begin{equation}\label{eq_sol_z}
\begin{aligned}
z(t)&=e^{A_{22}(t-t_{0})}z(t_{0})+\int_{t_{0}}^{t}e^{A_{22}(t-\tau)}A_{21}\Delta\omega_{r}(\tau)d\tau\\
&+\int_{t_{0}}^{t}e^{A_{22}(t-\tau)}B_{z}u_{s}(\tau)d\tau
\end{aligned}
\end{equation}
The mode where $\Delta\omega_{r}$ has the highest participation is the most relevant mode denoted by $\lambda_{r}$, and $\Delta\omega_{r}(\tau)$ can be expressed as $\Delta\omega_{r}(\tau)=c_{r}v_{r}e^{\lambda_{r}\tau}$ where $v_{r}$ is the corresponding eigenvector and $c_{r}$ is an arbitrary constant \cite{hector}. Since the electrical dynamics related to $A_{22}$ are faster than the electro-mechanical ones, the largest eigenvalue of $A_{22}$ is much smaller than $\lambda_{r}$. Thus, the natural response can be omitted. So, the first two terms in (\ref{eq_sol_z}) can be approximately calculated as \cite{hector}
\begin{equation}
\begin{aligned}
\label{eq_sol_integral_1}
e^{A_{22}(t-t_{0})}z(t_{0})+\int_{t_{0}}^{t}e^{A_{22}(t-\tau)}A_{21}\Delta\omega_{r}(\tau)d\tau\\
\approx (\lambda_{r}I-A_{22})^{-1}A_{21}\Delta\omega_{r}
\end{aligned}
\end{equation}
Since a Boolean control is considered, $u_{s}$ is constant. The second integral in (\ref{eq_sol_z}) can be computed as
\begin{align}
\label{eq_sol_integral_rest}
\int_{t_{0}}^{t}e^{A_{22}(t-\tau)}B_{z}u_{s}d\tau= (-A_{22})^{-1}B_{z}u_{s}
\end{align}
The response of the less relevant dynamics are expressed as
\begin{align}
\label{eq_z_solved}
&z\approx(\lambda_{r}I-A_{22})^{-1}A_{21}\Delta\omega_{r} + (-A_{22})^{-1}B_{z}u_{s}
\end{align}
Substituting (\ref{eq_z_solved}) into (\ref{eq_linear_ss_more}) and (\ref{eq_linear_ss_out}) yields the following reduced first-order model
\begin{align}
\label{eq_linear_ss_reduced}
\begin{split}
\Delta\dot{\omega}_{r}&=A_{\text{rd}}\Delta\omega_{r} + B_{\text{rd}}u_{s}\\
\Delta P_{g}&=C_{\text{rd}}\Delta\omega_{r} + D_{\text{rd}}u_{s}
\end{split}
\end{align}
where
\begin{align*}
\begin{split}
& A_{\text{rd}}=A_{11}+A_{12}(\lambda_{r}I-A_{22})^{-1}A_{21} \\
& C_{\text{rd}}=C_{r}+C_{z}(\lambda_{r}I-A_{22})^{-1}A_{21}\\
& B_{\text{rd}}=B_{r} + A_{12}(-A_{22})^{-1}B_{z}\\
& D_{\text{rd}}=D_{\text{sys}} + C_{z}(-A_{22})^{-1}B_{z}
\end{split}
\end{align*}

\subsection{Augmented Frequency Response Model}
Then, the AFR associated with the network in Fig. \ref{fig_Model_Network} can be expressed as follows
\begin{align}
\label{eq_AFR}
\begin{aligned}
&2H_{d}\Delta\dot{\omega}_{d}=\overline{f}(\Delta P_{m}-k_{d}\Delta P_{d}+k_{dw1}\Delta P_{g1}+k_{dw2}\Delta P_{g2})\\
&\tau_{d}\Delta\dot{P}_{m}=-\Delta P_{m}+\Delta P_{v}\\
&\tau_{g}\Delta\dot{P}_{v}= -\Delta P_{v}  - \Delta\omega_{d}/(\overline{f}R_{D})\\
&\Delta\dot{\omega}_{r1}=A_{\text{rd1}}\Delta\omega_{r1} + B_{\text{rd1}}u_{s1}\\
&\Delta\dot{\omega}_{r2}=A_{\text{rd2}}\Delta\omega_{r2} + B_{\text{rd2}}u_{s2}
\end{aligned}
\end{align}
where
\begin{align}
\begin{split}
&\Delta P_{g1}=C_{\text{rd1}}\Delta\omega_{r1} + D_{\text{rd1}}u_{s1}\\
&\Delta P_{g2}=C_{\text{rd2}}\Delta\omega_{r2} + D_{\text{rd2}}u_{s2}
\end{split}
\end{align}
Let $S_{d}$, $S_{w1}$ and $S_{w2}$ be the base of DG and WTG 1 and 2, respectively. Then, $k_{d}=1/S_{d}$, $k_{dw1}=S_{w1}/S_{d}$, and $k_{dw2}=S_{w2}/S_{d}$. The term $\Delta P_d$ is the worst-case contingency.

\section{MPC-based Control Synthesis with Temporal Logic Specifications}\label{sec_method}
\subsection{Overall Configuration}
\begin{figure}[h]
	\centering
	\includegraphics[scale=0.47]{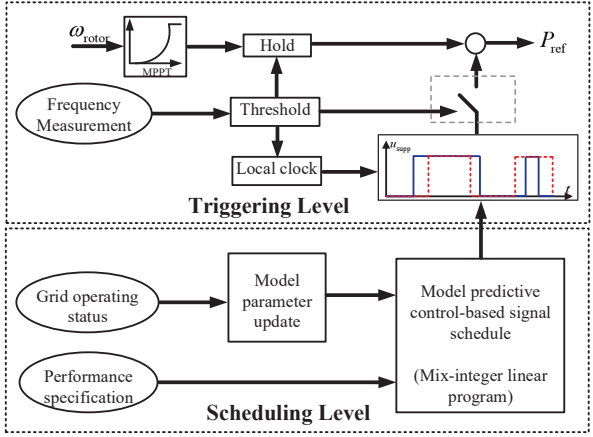}
	\caption{Overall configuration of synthesizing performance guaranteed controller.}
	\label{fig_overall}
\end{figure}

The overall configuration of the proposed control is illustrated in Fig. \ref{fig_overall}. The controller is configured into two levels, that is, the scheduling level and the triggering level. In the scheduling level, the grid operating status is acquired to update the parameters of the AFR model. The required performance specifications and up-to-date models are sent to the MPC-based signal scheduling program. The signals are Boolean with pre-specified magnitude. The signal scheduling problem is formulated as a MILP. Then, the supportive signals for WTGs can be pre-calculated under a worst-credit contingency.

The scheduled signals are sent to the triggering level, where the frequency is measured and compared to a pre-defined threshold to detect whether a severe contingency close to the worst-case one is happening. Once the supportive function is determined to be activated, a local clock is activated so that the scheduled signals are synchronized with the real time. And the synchronized signals are applied to the supplementary loop of the WTGs. It is worth mentioning that the initial condition in the MPC scheduling should be aligned with the threshold setting.

\subsection{MPC Formulation for Scheduling Level}
Define the state and input vectors as
\begin{equation}
\begin{aligned}
& x=[\Delta\omega_{d},\Delta P_{m},\Delta P_{v},\Delta\omega_{r1},\Delta\omega_{r2}]^{T}\\
& u=[u_{s1},u_{s2}]^{T}
\end{aligned}
\end{equation}
Then, the analytical model in (\ref{eq_AFR}) is discretized at a sample time of $t_{s}$ and expressed compactly as follows
\begin{align}
x(k+1)=A_{d}x(k)+B_{d1}u(k)+B_{d2}k_{d}\Delta P_{d}
\end{align}
Let the scheduling horizon be denoted as $k\in\mathcal{T}=[1,\cdots,T]$. First, the frequency deviation should not exceed a certain limit in any time, that is, 
\begin{align}
|x_{1}(k)|\leq \Delta f_{d,\text{lim}}\quad\forall k\in\mathcal{T}
\end{align}
Since the kinetic energy of WTGs will be transferred to active power to support the grid, the speed of WTGs will decrease from nominal values. This deviation is also desired to be limited for both WTGs
\begin{align}
|x_{i}(k)|\leq \Delta f_{w,\text{lim}}\quad\forall k\in\mathcal{T},i=4,5
\end{align}
The Boolean control signals for both WTGs can be presented using the following constraints
\begin{align}
u_{si}(k)=b_{i}(k)u_{C}\quad\forall k\in\mathcal{T},i=1,2
\end{align}
where $b_{i}$ is a binary variable indicating the status of the GS mode of WTG $i$, and $u_{C}$ is the fixed magnitude of the inputs. Finally, the frequency is required to satisfy the following TLS $\varphi$ to enhance the performance
\begin{align}
x_{1}(k)\vDash \varphi\quad\forall k\in\mathcal{T}
\end{align}
where
\begin{align}
\varphi=\square[ (|x_{1}(k)|\geq\Delta f_{c})\rightarrow
\lozenge_{[0,t_{a}]} \square(|x_{1}(k)|\leq \Delta f_{c})  ]
\end{align}
The above TLS states that whenever the frequency deviation is larger than $\Delta f_{c}$, then it should become less than $\Delta f_{c}$ within $t_a$ seconds.

The first objective is to minimize the control efforts. The total control effort can be represented as the summation of all binary variables as
\begin{align}
C_{U}=\sum_{i=1}^{2}\sum_{k=1}^{T}b_{i}(k)
\end{align}
In addition, the switching between on and off of the supportive modes should not be too frequent. Thus, a start-up cost is added as follows
\begin{align}
C_{SU}=\sum_{i=1}^{2}\sum_{k=2}^{T-1}b_{i}(k)[1-b_{i}(k-1)]
\end{align}
This nonlinear objective can be converted into a linear objective with constraints by introducing slack binary variable $z$ as follows
\begin{align}
C_{SU}^{\prime}=\sum_{i=1}^{2}\sum_{k=2}^{T-1}(b_{i}(k)-z_{i}(k))
\end{align}
and
\begin{equation}
\begin{aligned}
&z_{i}(k)\leq b_{i}(k),z_{i}(k)\leq b_{i}(k-1)\\
&z_{i}(k)\geq b_{i}(k)+b_{i}(k-1)-1\quad\forall k\in\mathcal{T},i=1,2
\end{aligned}
\end{equation}
The scheduling problem can be summarized as follows
\begin{equation}
\begin{aligned}
&\min\quad w_{1}C_{U}+w_{2}C_{SU}^{\prime}\\
&\text{s.t.}\quad\forall k\in\mathcal{T}\\
&x(k+1)=A_{d}x(k)+B_{d1}u(k)+B_{d2}k_{d}\Delta P_{d}\\
&|x_{1}(k)|\leq \Delta f_{d,\text{lim}}\\
&|x_{i}(k)|\leq \Delta f_{w,\text{lim}}\quad i=4,5\\
&u_{i}(k)=b_{i}(k)u_{C}\quad i=1,2\\
&z_{i}(k)\leq b_{i}(k),z_{i}(k)\leq b_{i}(k-1)\quad i=1,2\\
&z_{i}(k)\geq b_{i}(k)+b_{i}(k-1)-1\quad i=1,2\\
&x_{1}(k)\vDash \varphi\\
&\varphi=\square[ (|x_{1}(k)|\geq\Delta f_{c})\rightarrow
\lozenge_{[0,t_{a}]} \square(|x_{1}(k)|\leq \Delta f_{c})  ]
\end{aligned}
\end{equation}
where $w_{1}$ and $w_{2}$ are positive weighing factors. The TLS can be encoded into a MILP using the toolbox BluSTL \cite{donze2015blustl}. Then, the overall problem is converted into a MILP, written in the format of Yalmip \cite{yalmip} and solved by efficient solvers Mosek \cite{mosek} and Gurobi.

\subsection{Results and Simulation Verification}\label{sec_sub_sim}
The rated powers of DG and WTG are assumed to be 2 MW and 1 MW, respectively. The operating conditions of the WTGs and their corresponding first-order model are given as follows
\begin{equation*}
\begin{split}
&v_{\text{wind}}=10\text{ [m/s]},P_{gi}=0.8,Q_{gi}=0,v_{dsi}=0,v_{qsi}=1\\
&A_{\text{rd}i}=-0.2771, B_{\text{rd}i}=2.5741, C_{\text{rd}i}=0.2550, D_{\text{rd}i}=-2.3343
\end{split}
\end{equation*}
for $i=1,2$. The parameters associated with the DG are given as follows
\begin{equation*}
\begin{split}
& H_{d}=4,\tau_{d}=0.1,\tau_{g}=0.5
\end{split}
\end{equation*}
The base and scaling factors are
\begin{equation*}
\begin{split}
& S_{d}=5\text{ [MVA]},S_{w}=1.11\text{ [MVA]},k_{d}=0.2,k_{wd}=0.22
\end{split}
\end{equation*}
The parameters in the MILP are given as follows
\begin{equation*}
\begin{split}
& t_{s}=0.02\text{ [s]},T=4\text{ [s]},P_d=0.7\text{ [MW]},w_{1}=1,w_{2}=10\\
& \Delta f_{d,\text{lim}}=0.5\text{ [Hz]},\Delta f_{w,\text{lim}}=2\text{ [Hz]},u_{C}=-0.05\\
& f_{c}=0.45+\varepsilon\text{ [Hz]},t_{a}=1\text{ [s]}
\end{split}
\end{equation*}
Based on the given parameters, it is required that the frequency deviation to be limited within 0.5 Hz. Moreover, whenever the frequency deviation is larger than 0.45 Hz, it should be restored back to 0.45 Hz within one second. Since there exists certain mismatches between the AFR and the full nonlinear model, the term $\varepsilon$ is introduced to tighten the specification such that the nonlinear response can satisfy the original specification as well.

\begin{figure}[htbp!]
	\centering
	\includegraphics[scale=0.5]{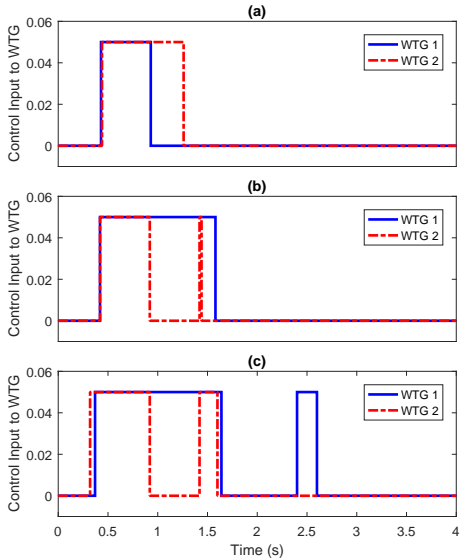}
	\vspace{-0.15in}
	\caption{Scheduled control signals for WTGs. (a) Without TLS. (b) With TLS. (c) With TLS and a robust margin.}
	\label{fig_Control}
\end{figure}

\begin{figure}[htbp!]
	\centering
	\includegraphics[scale=0.5]{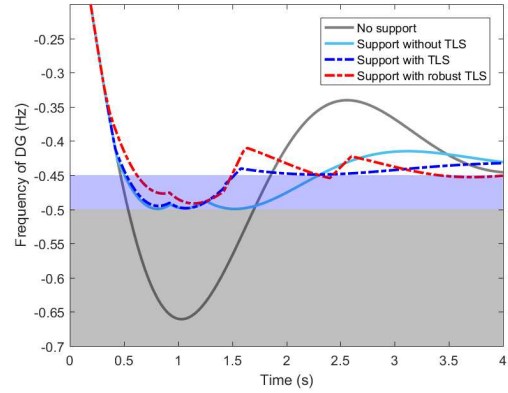}
	\vspace{-0.05in}
	\caption{Frequencies of DG under different cases simulated using the AFR model.}
	\label{fig_Freq_linear}
\end{figure}

Three cases are considered. In the first case, the TLS is removed. In the second case, the TLS is considered with the compensating factor $\varepsilon=0$. In the third case, the compensating factor $\varepsilon$ is set to be $-0.015$ Hz. The scheduled inputs of these three cases are plotted in Fig. \ref{fig_Control}. The DG frequencies under these cases from the AFR are shown in Fig. \ref{fig_Freq_linear}. As shown, with more constraints, the WTGs are required to operate at the GS mode for larger time durations. The responses from the AFR model strictly satisfy all control specifications with minimum control efforts required.

\begin{figure}[htbp!]
	\centering
	\includegraphics[scale=0.5]{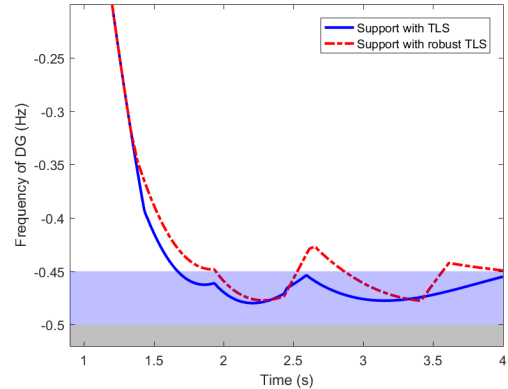}
	\vspace{-0.05in}
	\caption{Frequencies of DG under different cases simulated using the full nonlinear model in Simulink.}
	\label{fig_Freq_non}
\end{figure}

\begin{figure}[htbp!]
	\centering
	\includegraphics[scale=0.5]{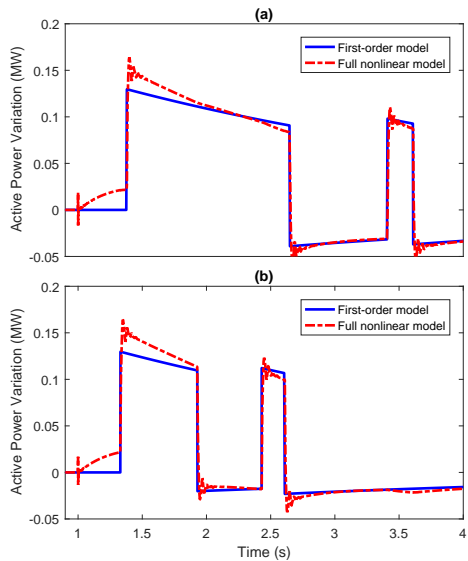}
	\vspace{-0.15in}
	\caption{Active power variations from the first-order and full nonlinear model. (a) WTG 1. (b) WTG 2.}
	\label{fig_WTG_non}
\end{figure}

The scheduled inputs of Case 2 and 3 are applied to the nonlinear model. The corresponding frequencies of DG are shown in Fig. \ref{fig_Freq_non}. The DG frequency in Case 2 does not satisfy the TLS. This is because of the error induced by the model reduction of WTGs. The active power variations associated with the support signals in Case 3 are shown in Fig. \ref{fig_WTG_non}. As shown, although the first-order models have successfully captured the active power dynamics with good accuracy, there are still mismatches in the response. These tiny mismatches, however, falsify the TLS, the satisfaction of which requires higher level precision. Thus, the response mismatches need to be compensated. The most convenient approach is to impose more strict specifications, that is, the introduction of the robust factor $\varepsilon$, such that the output could satisfy the original specifications at the cost of introducing certain levels of conservatism. The red dash plot in Fig. \ref{fig_Freq_non} indicates that this robust factor could generate a stronger control effort so that the specifications are satisfied. It is also worth mentioning that in the nonlinear verification, the TLS is a bit conservative because the AFR model is not able to capture the weak inertial responses from the DFIG-based WTGs. 

\section{Conclusions and Future Works} \label{sec_conclusion}
In this paper, a MPC-based control synthesis methodology is proposed that enables the realization of the TLSs. The controller schedules ahead a series of Boolean control signals to synthesize the GS mode of WTGs by solving the MPC problem, where the frequency response predicted by the AFR model satisfies the defined specifications under a worst-case contingency. The proposed control is verified on the full nonlinear model in Simulink. A robust factor is introduced to compensate the model reduction error such that the nonlinear response satisfies the TLS. The future work will be devoted to the development of a hierarchical configuration for larger-scale systems. Meanwhile, a systematic approach will be studied to attain a good trade-off between error compensation and conservatism.

\section{Acknowledgment}
Research sponsored by the Laboratory Directed Research and Development Program of Oak Ridge National Laboratory (ORNL), managed by UT-Battelle, LLC for the U.S. Department of Energy under Contract No. DE-AC05-00OR22725. The United States Government retains and the publisher, by accepting the article for publication, acknowledges that the United States Government retains a non-exclusive, paidup, irrevocable, world-wide license to publish or reproduce the published form of this manuscript, or allow others to do so, for United States Government purposes. The Department of Energy will provide public access to these results of federally sponsored research in accordance with the DOE Public Access Plan (http://energy.gov/downloads/doe-public-access-plan).

\bibliography{IEEEabrv_zyc,Ref_LTL}
\bibliographystyle{IEEEtran}

\end{document}